\documentstyle[aps,psfig,twocolumn]{revtex}
\newcommand{\s}{\mathrm}

\newcommand{\ra}{\rightarrow}

\newcommand{\be}{\begin{equation}}
\newcommand{\ee}{\end{equation}}

\newcommand{\bea}{\begin{eqnarray}}
\newcommand{\eea}{\end{eqnarray}}
\newcommand{\bef}{\begin{figure}}
\newcommand{\eef}{\end{figure}}

\newcommand{\lgl}{\langle}
\newcommand{\rgl}{\rangle}

\tightenlines
\begin{document}
\draft
\title{ Spectral Red-Shift Versus Broadening from
Photon and Dilepton Spectra}
\vskip 0.2 in
\author{Jan-e Alam$^{1,3}$, Pradip Roy$^{2}$, Sourav Sarkar$^{3}$ and Bikash 
Sinha$^{2,3}$}
\address{$^1$~Physics Department, University of Tokyo, Tokyo 113-0033, Japan}
\address{$^2$~Saha Institute of Nuclear Physics, 
1/AF Bidhannagar, Kolkata 700064}
\address{$^3$~Variable Energy Cyclotron Centre, 1/AF Bidhannagar, Kolkata 700064}
\maketitle
\begin{abstract}
We estimate the photon and dilepton emission rates from hot hadronic 
matter with in-medium spectral shift and broadening of vector mesons. 
It is observed that both the WA98 photon data and CERES/NA45 dilepton 
data  can be well reproduced with similar initial conditions. 
The freeze-out condition has been constrained by the transverse 
mass spectra of pions and protons measured by the NA49 collaboration.
We argue that simultaneous measurement of the $p_T$ spectra of single 
photons as well as invariant mass distribution of dileptons is crucial 
to understand the in-medium spectral function of the vector mesons. 

\end{abstract}

\narrowtext
\section*{1. Introduction}

The study of the behaviour of hadrons
at finite temperature and/or high baryon density has 
attracted substantial theoretical and experimental attention
in recent times. 
The theoretical motivation stems from the expectation that 
broken chiral symmetry of QCD might be restored at high 
temperature and/or density, at least partially, and this has nontrivial
effects on the spectra of low lying hadronic states.
Again, these studies are of considerable importance in connection with the
experimental detection and study of quark gluon plasma 
(QGP) in the ultra-relativistic collisions of heavy ions.
This is because disentangling the signals from quark matter would require
a precise estimate of emissions from hadronic sources which constitutes an
overwhelming background in such experiments. 
Now, the properties of short-lived resonances like
the $\rho$ and $\omega$ mesons which are likely to decay
within the fireball can only be studied through
deep probes such as photons and lepton pairs which are essentially
electromagnetically interacting and thus probe the entire
space-time volume of the collision~\cite{1st}. 
The medium modifications of low mass vector mesons has been the subject 
of numerous theoretical investigations culminating broadly into
the following two types of predictions. QCD based models predict a 
downward shift in the mass, which we will mention as spectral red-shift and a 
rescattering scenario giving rise to an enhanced width which we have denoted by
spectral broadening, the mass remaining largely unchanged. Our aim 
in this work is to calculate the electromagnetic spectra considering
various plausible scenarios and compare with the WA98~\cite{2nd}
photon and CERES/NA45~\cite{ceres}
dilepton spectra in order to comment on the relative importance of
the two scenarios mentioned above.

Our analysis will proceed as follows. In the next section we will
define the effective mass and width that have been used in the calculation
and define the scenarios of medium effects considered. At the onset we
emphasize that our main objective is to see the effect of these
contrasting scenarios on the electromagnetic spectra and not on the
details of how the forms of the effective mass and the width have been arrived 
at. Consequently, we will use phenomenological arguments to justify
our choice of the parametrisations for these quantities. In section 3
we will discuss the initial conditions in considerable detail. In the
relevant transverse momentum region of the photon spectra
the main sources of photons are the hard pQCD processes as well as
the thermal emissions. Consequently, the number of participants as well
as the initial temperature are the relevant initial conditions which have
been estimated. For the low mass dilepton spectra, the contribution
from Dalitz decays as well as thermal emissions are important.
Since photons and dileptons are emitted at all stages of the collision, the
space-time evolution plays a very significant role in estimating the
total yield. We have used two different approaches to study this aspect;
relativistic hydrodynamics and transport. As we shall see, 
medium effects do play a very important 
role in the equation of state and consequently the cooling profile.
These aspects have been dealt with in section 4.
The transverse mass $(m_T)$ spectra of pions and protons obtained 
at SPS have been evaluated and compared with data from NA49~\cite{na49}
in section 5 
in order to estimate the freeze-out condition which goes as an input
into the hydrodynamic calculations. With all these inputs we evaluate
the photon and dilepton spectra in sections 6 and 7 respectively.
section 8 contains a summary and discussions.

\section*{2. Effective Mass and Width of Hadrons}

It has been emphasized that the properties
of hadrons will be modified due to its interaction with 
the particles in the thermal bath 
and such modifications will be reflected
in the dilepton and photon spectra
emitted from the system (see Refs.~\cite{geb,12th,13th,14th,15th} for review).
Broadly two types of medium modifications are expected:
shift in the pole position and/or broadening
of the spectral function.
As discussed earlier, our main objective is to comment on
the nature of medium effects through the study of 
WA98 photon and CERES dilepton data. Accordingly we consider
the following scenarios: (I) the system is formed
in the hadronic phase with the hadronic masses (except
pseudoscalars) approaching zero
near the critical temperature according to the universal scaling
law ~\cite{geb,scale}. The in-medium masses of the hadrons
and the decay width of $\rho$ are taken as
\bea
m_H^\ast/m_H&=&(1-T^2/T_c^2)^{\lambda}\nonumber\\
\Gamma_\rho&=&\frac{g_{\rho\pi\pi}^2}{48\pi}m_\rho^*
(1-4m_\pi^2/m_\rho^{*2})^{3/2}\left(1+2f_{BE}\right)
\label{emass}
\eea
where $f_{BE}$ is the Bose-Einstein distribution. For the
width of $\omega$ see Ref.~\cite{15th}.
(II) the width of the vector meson ($\rho$)
increasing with temperature as,
\bea
\Gamma_{\rho}^\ast &=& \Gamma_\rho/(1-T^2/T_c^2),\nonumber\\
m_H^\ast&=&\,m_H
\label{ewidth}
\eea
and the masses remain constant at their vacuum values.
The value of $\lambda$ in eq.~\ref{emass} is 1/6 and 1/2 for Brown-Rho
and Nambu scaling~\cite{geb} respectively. The present experimental
statistics can not differentiate among different values
of $\lambda$. Therefore, we leave it as a parameter here.
The scenario II derives motivation from
the results of chiral models according to which the mass of the $\rho$
does not change to order $T^2$ but the leading order $T$-dependence of
the pion decay constant results in the $\rho$ decay width having
the form described by Eq.~\ref{ewidth}.
We have also considered a third case, (III) in which both
the masses and widths are maintained at the vacuum values.
In eqs.~\ref{emass} and \ref{ewidth}, $T_c$  is the
critical temperature of chiral symmetry restoration
where the quark condensate $\langle \bar q q\rangle$ and
consequently the mass of the hadrons tend to zero.

Let us study the in-medium effects
on the thermal phase space factor 
through the $\rho$ meson as it plays the most important role
for the electromagnetic probes.
In Fig.~\ref{fig1} we
display the change in the density of thermal $\rho$
as a function of temperature for scenarios (I), (II)
and (III) by solid, dotted and long-dashed lines respectively.
It is clear that the density of $\rho$ is  very sensitive
to the spectral shift of $\rho$ mass due to
Boltzmann enhancement but rather insensitive to the
change of width.
The reason for small change in the density of $\rho$
due to the in-medium broadening can
be understood from the following arguments.
The density of an unstable vector meson (say, $\rho$) in a
thermal bath can be written as~\cite{19th},
\be
\frac{dN}{d^3kd^3xds}=\frac{g}{(2\pi)^3}\frac{1}{e^{\sqrt{k^2+s}/T}-1}P(s)
\label{uns}
\ee
where $g$ is the statistical degeneracy of the particle and $P(s)$ is
the spectral function,
\be
P(s)=\frac{1}{\pi}\frac{{\s {Im}}\,\Pi}{(s-m_\rho^2-{\s {Re}}\,\Pi)^2
+({\s {Im}}\,\Pi)^2}
\label{spfn}
\ee
${\s {Im}}\,\Pi$ (${\s {Re}}\,\Pi$) is the imaginary (real) part of the
$\rho$ self energy.
Eqs.~\ref{uns} and ~\ref{spfn}
indicate that the density of particles (vector mesons) in a thermal
bath is given by the Bose-Einstein distribution weighted by the Breit-Wigner
function, which gets maximum weight from the
value of $s=m_\rho^2+{\s {Re}}\,\Pi$,
the contribution
from either side of the maximum being averaged out.
Therefore, the results become sensitive to the
value of $s=m_\rho^{2*}=m_\rho^2+{\s {Re}\Pi}$ and not to
he width of the spectral distribution.
Note that $P(s)\,$ tends to $\,\delta(s-m_\rho^2-{\s {Re}}\Pi)$
as ${\s {Im}}\,\Pi\,\ra\,0$,
corresponding to a stable particle.  (Here,
$\Pi$ is proportional to the trace of the self energy tensor
$\Pi_{\mu\nu}$ of the $\rho$).

\bef
\centerline{\psfig{figure=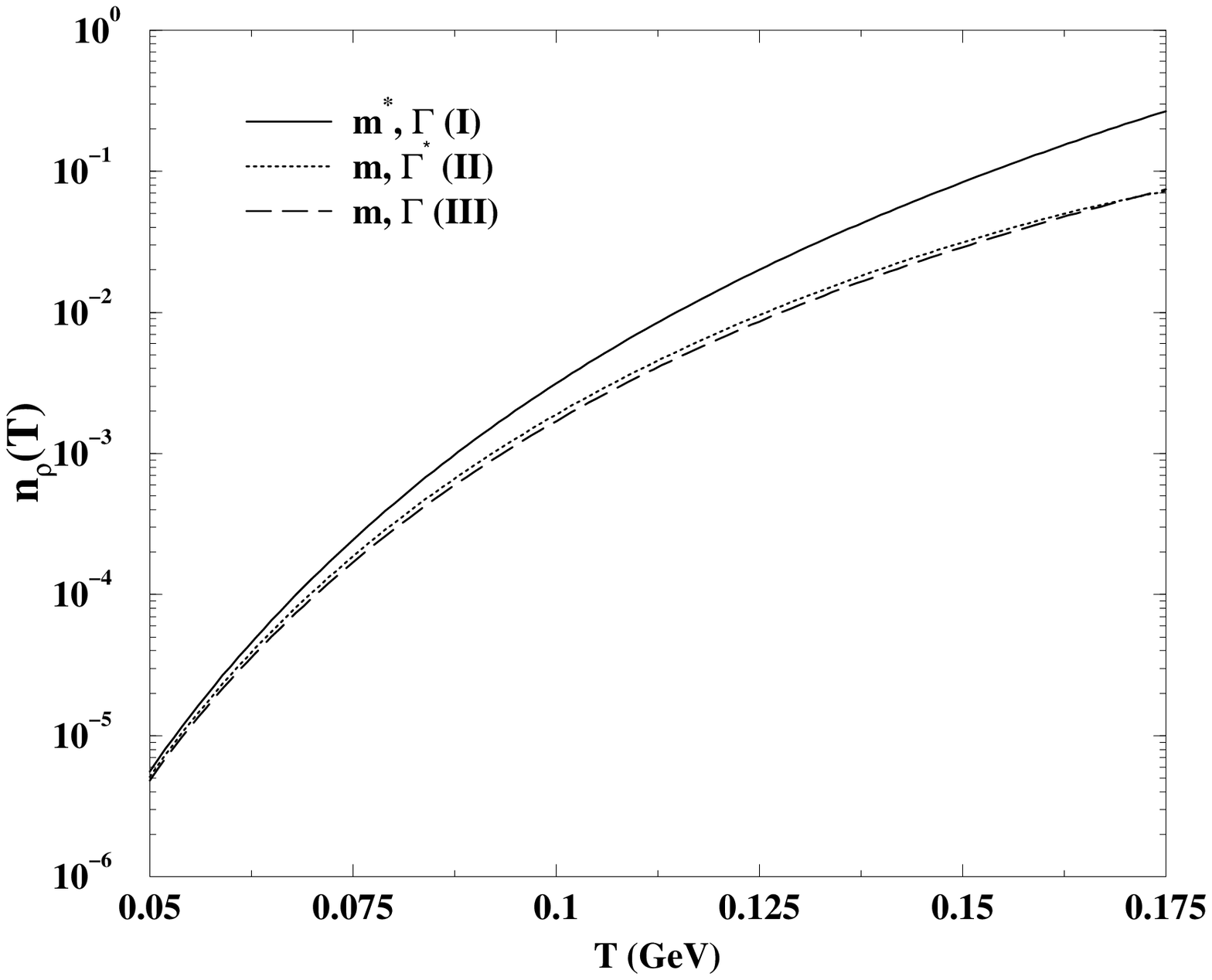,height=5.5cm,width=8cm}}
\caption{~Solid (long dashed) line
represent density of thermal $\rho$ as a function of temperature
with in-medium mass and vacuum width (vacuum mass and width).
Dotted line represents the density with vacuum mass and
in-medium width.
}
\label{fig1}
\eef

\section*{3. Initial Conditions}

Here we will discuss the parameters which enter as inputs to
the evaluation of electromagnetic and hadronic spectra 
from relativistic heavy ion collisions corresponding to
different scenarios considered.
We begin with the quantity $n(b)$ which is the
average number of nucleon-nucleon collisions in the collision of two nuclei
of mass numbers $A$ and $B$ at an impact parameter $b$.
This is given by~\cite{cywbook},
\be
n(b)=ABT_{AB}(b)\sigma_{in}
\label{bden}
\ee
where $T_{AB}(b)$ is the nuclear overlap integral evaluated
from the following expression,
\be
T_{AB}(\vec{b})=\int d^2sT_A(\vec{s})T_B(\vec{b}-\vec{s}),
\ee
and the thickness function, $T_A$ is defined as
$T_A(\vec b)=\int dz \rho_A(\vec b,z)$. The nuclear density,
$\rho_A(\vec b, z)$ is parametrized by Wood-Saxon type profile
function~\cite{kje},
\be
\rho_A(r)=\rho_0\frac{1+\omega\,r^2/R_A^2}{1+
e^{(r-R_A)/z}}
\ee
where $\omega=0$, $z=0.549$ for the Pb
nucleus and $R_A=1.2A^{1/3}$ and
the central density, $\rho_0$ is determined by the normalization condition
$\int d^3r\rho_A(r)=1$.
In Fig.~\ref{fig0} $n(b)$ is shown as a function of the
impact parameter, $b$. The most central event in WA98
corresponds to $b\,\sim\, 3.2$ fm/c. At this value of 
the impact parameter $n(b)\sim 660$. 

\bef
\centerline{\psfig{figure=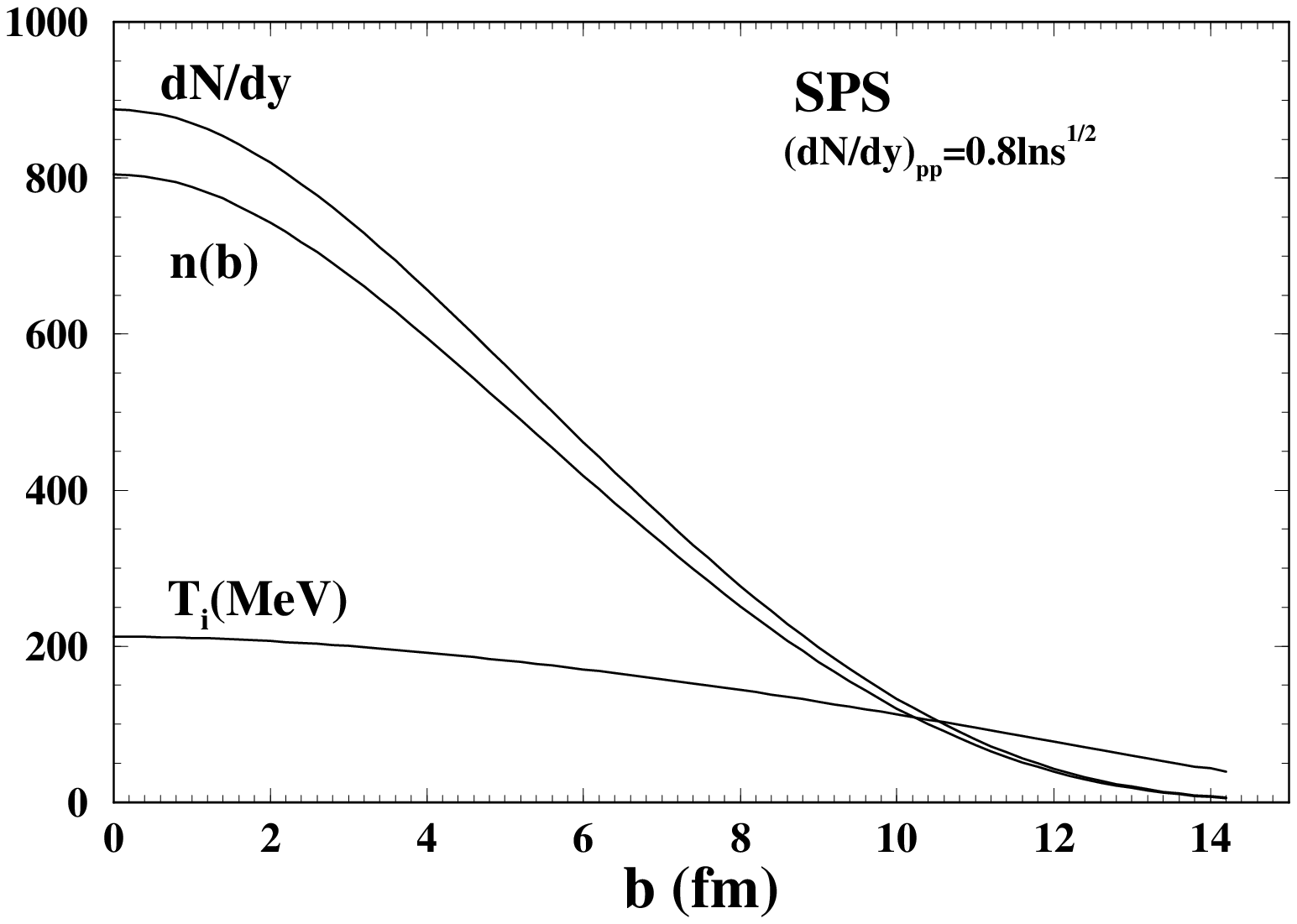,height=5.5cm,width=8cm}}
\caption{~Hadron multiplicity, effective number of nucleon-nucleon
collisions and initial temperature as a function of impact
parameter calculated by using Glauber model for
nucleus - nucleus collisions (see text).
}
\label{fig0}
\eef

The hadron multiplicity in A + B collisions
at an impact parameter $b$ can then be evaluated in
terms of the multiplicity in $pp$ collisions as~\cite{cywbook},
\be
\frac{dN}{dy}(b)=\frac{n(b)}{1+\delta(A^{1/3}+B^{1/3})}\frac{dN_{pp}}{dy},
\ee
where $dN_{pp}/dy=0.8\ln\sqrt{s}$ is the hadron multiplicity
for nucleon-nucleon collisions at mid-rapidity~\cite{UH}.
The value of $dN/dy$ is 725 at $b\sim 3.2$ fm,
$\sqrt{s}=17.3$ GeV and $\delta=0.09$ (see Fig.\ref{fig0}).
The term $1+\delta(A^{1/3}+B^{1/3})$
takes into account the energy degradation of the participating
nucleons in the nuclear environment. For isentropic expansion
the initial temperature ($T_i$) is related to the hadron
multiplicity as,
\be
T_i^3(b)=\frac{2\pi^4}{45\xi(3)}\frac{1}{\pi R_A^2\tau_i 4a_k}\,
\frac{dN}{dy}(b)
\ee
$a_k=\pi^2g_k/90$ is determined by the statistical degeneracy ($g_k$)
of the system formed after the collision. In case of a deconfined initial
state, $g_k=37$ considering a non-interacting gas of gluons and quarks of two
flavours. 
For a hadron gas composed of $\pi$, $\rho$,
$\omega$, $\eta$, $a_1$ and nucleons
the effective degeneracy has a value $\sim 30$ near the critical
temperature where the hadronic masses go to zero according
to the universal scaling scenario~\cite{geb}.
The increase in the degeneracy originates
from the heavier hadrons going to a massless situation~( see also \cite{VK}).
A value of $g_{eff}\sim 30$ can also be realized from the lattice
data~\cite{FK,MA} near the  phase transition point by using the
relation $s/T^3=4\pi^2 g_{eff}/90$, where $s$ is
the entropy density.
The latter is calculated using the relation, $s=(\epsilon+P)/T$ and
contains the effect of the in-medium masses of the
constituent hadrons (for details see~\cite{15th}).
For $\tau_i=1$ fm/c~\cite{jdb} and $g_k\sim 30$,
we get an initial temperature $T_i(b\sim 3.2$ fm)
$\sim 200$ MeV (see Fig.~\ref{fig0}).
Note that the variation of $T_i(b)$ as a function of
the impact parameter is very slow because $T_i(b)\sim
[dN/dy(b)]^{1/3}$. 
In an attempt to understand the lattice data in
terms of the effective fields~\cite{HAB}
the effective degrees of freedom in hadrons
was limited to $\sim 24$ which coincides with the number of fundamental
degrees of freedom for two quark flavors.

\section*{4. Space-Time Evolution and Equation of State}

Next we study the sensitivity of the results
on the space-time evolution. In order to do that
we solve the (3+1) dimensional hydrodynamic equations with
initial energy density~\cite{hvg},
\be
\epsilon(\tau_i,r) =\frac{\epsilon_0}{e^{(r-R_A)/\delta}+1}
\label{enerfile}
\ee
and initial velocity profile,
$v_r=v_0\left(r/R_A\right)^\alpha$
which has been
successfully used to study transverse momentum spectra of hadrons
~\cite{pbm,uh} and photons~\cite{5th,6th,7th}.
For our numerical calculations we choose $\alpha=1$
and $v_0=0$ at the initial time.

The equation of state (EOS) used here to solve the
hydrodynamic equations is  evaluated
as in Ref.~\cite{15th}. The temperature
dependence of the mass enters into the EOS through
the effective statistical degeneracy ($g_{eff}$).
In Fig.~\ref{fig4} the temperature variation of the effective
degeneracy obtained from different models for
the hadronic interactions (see ~\cite{15th} for details)
is compared with the lattice QCD calculations~\cite{fk}.
The co-efficient of $\epsilon/T^4$ differs
from the effective degeneracy by a factor of $\pi^2/30$.
The lattice data seems to be well reproduced by the
universal scaling scenario of mass reduction given
by eq.~\ref{emass}.
In this case the velocity of sound ($c_s$) is given by
$c_s^{-2}=Tds/sdT=[(T/g_{eff})(dg_{eff}/dT)+3]$~\cite{ss}.
Clearly, the value of $c_s$ is less than its value
corresponding to ideal fluid case ($c_s^{ideal}=1/\sqrt{3}$).
This affects the space time evolution of the system
non-trivially. In solving
the hydrodynamic equations we have used the equation
of state which contains the in-medium shift
of the hadronic spectral functions as described above.

The integration over the space time history has
also been performed by taking the temperature profile
from the transport model where the temperature varies with
time as follows:
\be
T(t)=(T_i-T_\infty)e^{-t/\tau}+T_\infty
\label{cool}
\ee
The cooling law given above is the
parametrization of the  results of~\cite{18th},
used in several articles in the literature~\cite{rcw}
(see also the review~\cite{14th} and references therein).
The calculation is performed with the following values of parameters:
$T_i=200$ MeV, $T_\infty=120$ MeV, $\tau=8$ fm/c.

\bef
\centerline{\psfig{figure=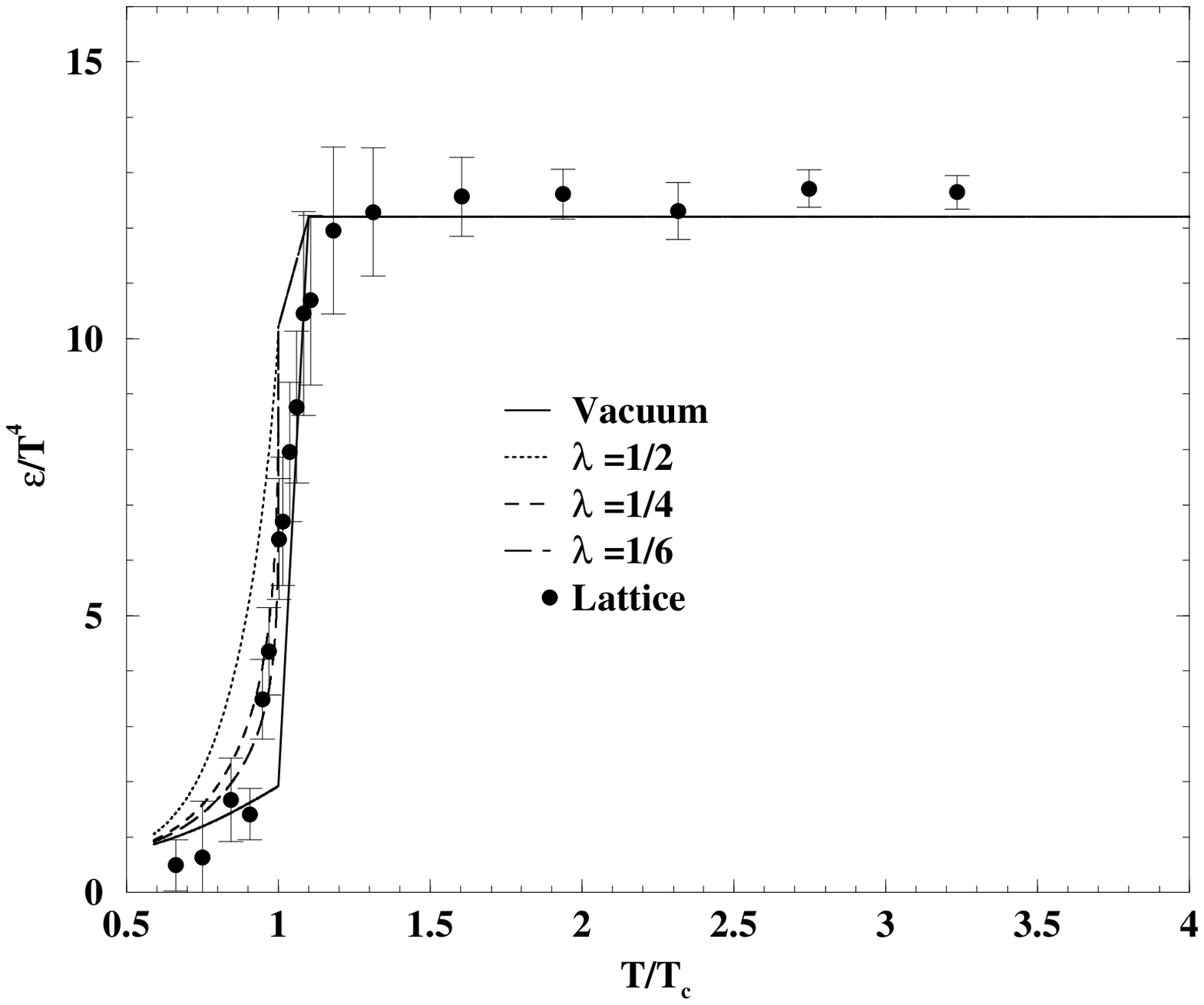,height=7.5cm,width=8cm}}
\caption{The energy density $\epsilon$ in
the unit of $T^4$ for the equation of state
for various models of hadronic interactions
is plotted as function of temperature ($T$) in the unit
of the critical temperature, $T_c$.
The filled circle denotes the lattice
results~{\protect\cite{fk}}.
The universal scaling scenario
seems to describe the lattice results quite well near the $T_c$.
For $T>T_c$ the bag model equation of
state has been used.
}
\label{fig4}
\eef

\section*{5. Hadronic Spectra and Freeze-out Temperature}

Having fixed the initial temperature and the equation of state
we now need to fix another important parameter, the freeze-out
temperature $T_F$ where the hydrodynamic evolution should terminate.
At this stage the mean free path of the constituents begin
to exceed the size of the system and particles start free streaming
to the detector.
For this purpose we will consider the $m_T(=\sqrt{p_T^2+m^2}$) 
spectra~\cite{ruuskanen,blaizot} of pions and
protons measured in Pb + Pb collisions at SPS energies by the 
NA49 collaboration~\cite{na49}.
We will assume that pions and nucleons are in thermal equilibrium
throughout the evolution until they freeze-out
at a common temperature $T_F$.
Negative hadrons and positive
minus the negative are treated as pions and protons respectively.
The $\pi^-$ and $K^-$ from the decays
$\rho$ and $\phi$ are about $5\%$ of the direct pions
at $p_T\sim 500$ MeV and are therefore neglected here.
In fig.~\ref{hadron1} NA49 pion spectra is compared  with
the hadronic initial states with $T_i\sim 200$ MeV and
$T_F=$ 100, 120 and 140 MeV. We find a reasonable agreement for $T_F=120$ MeV
as far as the slope is concerned. Calculations are done with all
the scenarios I, II and III. Due to reasons explained in section 2,
the results with scenario II are indistinguishable from that of scenario III.
We arrive at similar conclusions from the analysis of the proton spectra 
as shown in fig.~\ref{hadron2}. 

\bef
\centerline{\psfig{figure=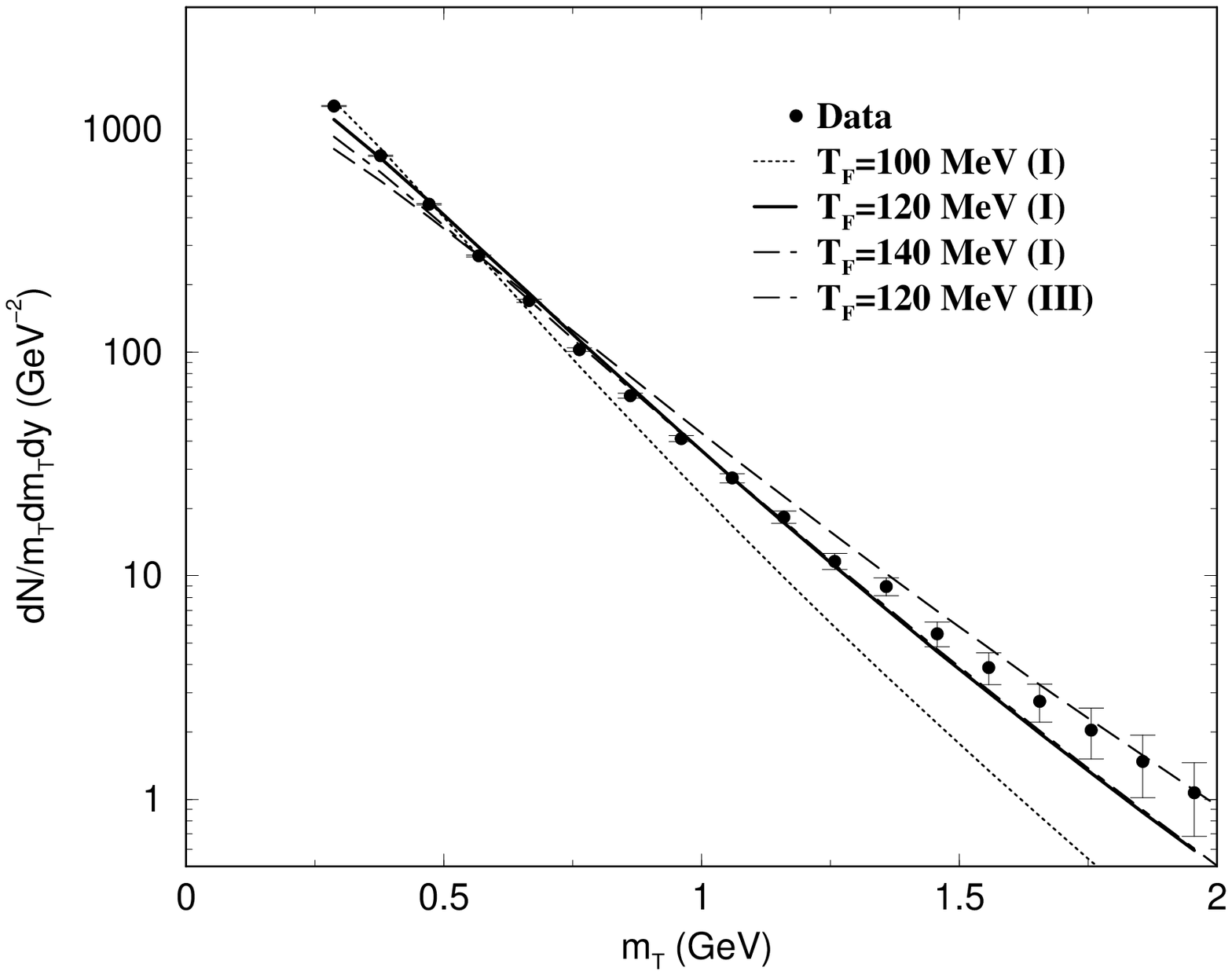,height=5.5cm,width=8cm}}
\caption{The $m_T$ distribution of pions for Pb + Pb collisions
at CERN SPS energies.
}
\label{hadron1}
\eef

\bef
\centerline{\psfig{figure=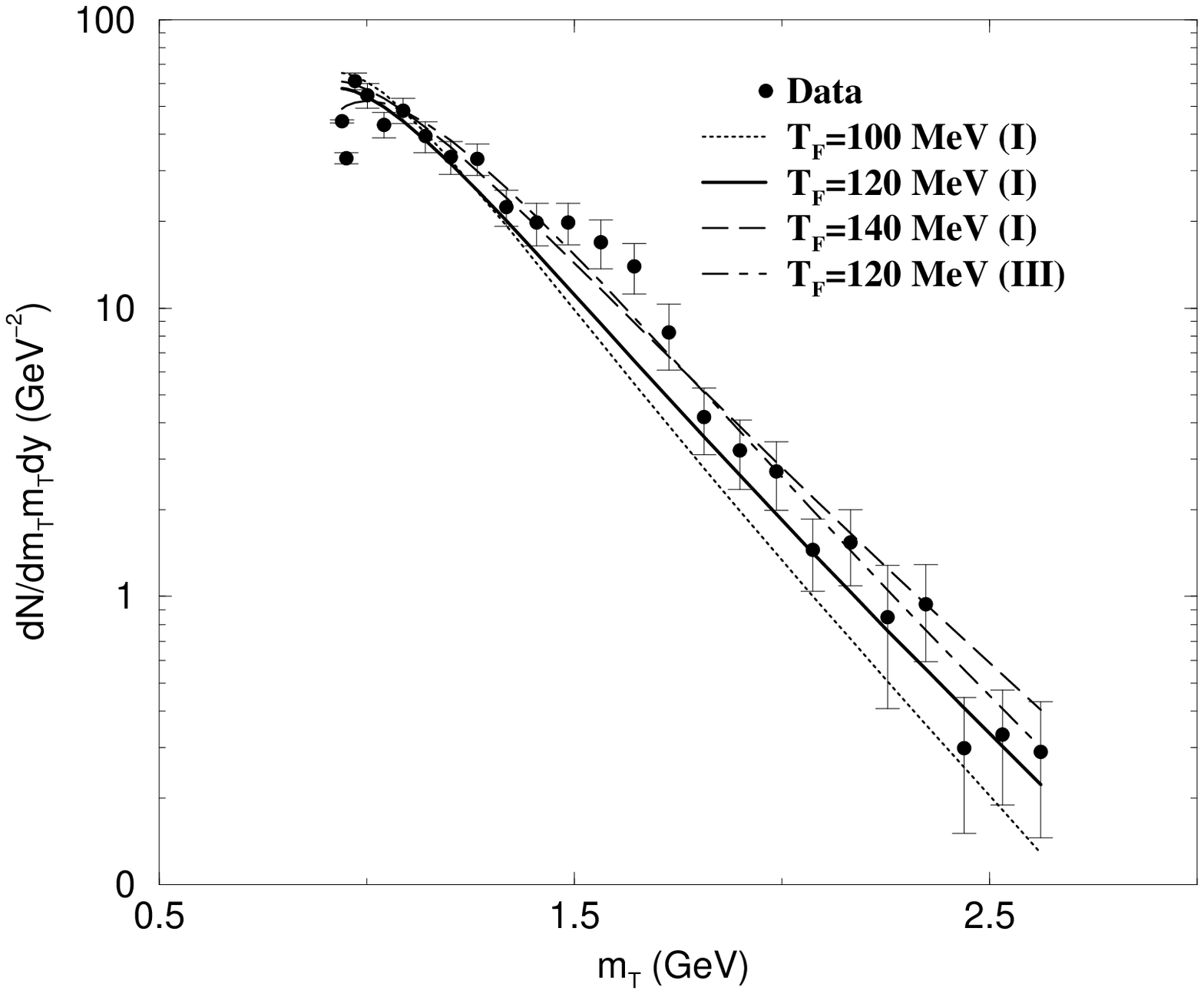,height=5.5cm,width=8cm}}
\caption{The $m_T$ distribution of protons for Pb + Pb collisions
at CERN SPS energies.
}
\label{hadron2}
\eef


\section*{6. Photon Spectra}

We start with the single photon data of
WA98 collaboration which has initiated 
considerable theoretical activities~\cite{5th,6th,7th,3rd,4th,8th}.
For the analysis of WA98 photon spectra, we consider
prompt photons resulting from the hard collisions of the 
partons in the colliding nuclei and the thermal
photons from a hot hadronic gas with possible in-medium
modifications.  Photons from hadronic decays 
($\pi^0\,\ra\,\gamma\gamma$, $\eta\,\ra\,\gamma\gamma$ etc.)
are already subtracted from the data~\cite{2nd} and
hence we need not consider them here. We begin our discussions
with the prompt photon. 
The prompt photon yield for nucleus-nucleus collision is given by,
\be
E\frac{dN}{d^3p}=n(b)\frac{1}{\sigma_{in}}\,\,E\frac{d\sigma_{pp}}{d^3p}
\ee
where $n(b)$ is the average number of nucleon-nucleon 
collisions at an impact parameter $b$, 
and $\sigma_{in}$ (=30 mb, taken from Ref.~\cite{2nd})
is the $p-p$  inelastic cross section. 
The prompt photon contributions have been evaluated
with possible intrinsic transverse motion of the partons~\cite{4th,9th}
inside the nucleon and  multiplied by a $K$-factor $\sim 2$,
to account for the higher order effects. The CTEQ(5M) parton
distributions~\cite{cteq} have been used for evaluating hard photons.
The relevant value of $\sqrt{s}$, energy in the centre of mass
for WA98 experiment is 17.3 GeV. No experimental data on hard
photons exist at this energy. Therefore, the ``data'' at $\sqrt{s}=17.3$
GeV is obtained from the data at $\sqrt{s}=19.4$ GeV of the E704
collaboration ~\cite{10th}
by using the scaling relation: 
 $Ed\sigma/d^3p_\gamma\mid_{h_1+h_2\,\ra\,C+\gamma}=f(x_T=2p_T/\sqrt{s})/s^2$,
for the hadronic process, $h_1+h_2\ra C+\gamma$~\cite{TF}.
This scaling is valid in the naive parton model. However,
such scaling may be spoiled in perturbative QCD due to
the reasons, among others, 
the momentum dependence of the 
strong coupling, $\alpha_s$ and from the scaling violation
of structure functions, resulting  
in faster decrease of the cross section than
$1/s^2$~\cite{9th}. Therefore, the data at $\sqrt{s}=17.3$ GeV
obtained by using the above scaling gives a conservative
estimate of the prompt photon contributions.  
We have seen that the effects of the nuclear shadowing
in parton distributions on the prompt photons are
negligibly small at SPS but it is important at RHIC
and LHC energies~\cite{pkr}. This is because the
value of $x(\sim 2p_T/\sqrt{s})$ 
at SPS is not small enough for the shadowing
effects to be important.

To evaluate the photon emission rate from a hadronic gas
we  model the system as consisting of $\pi$, $\rho$, $\omega$
$\eta$ and $a_1$~\cite{16th,17th}. The relevant vertices for the reactions
$\pi\,\pi\,\ra\,\rho\,\gamma$ and $\pi\,\rho\,\ra\,\pi\,\gamma$
and the decay $\rho\,\ra\,\pi\,\pi\,\gamma$
are obtained from the following Lagrangian:
\be
{\cal L} = -g_{\rho \pi \pi}{\vec {\rho}}^{\mu}\cdot
({\vec \pi}\times\partial_{\mu}{\vec \pi}) - eJ^{\mu}A_{\mu} + \frac{e}{2}
F^{\mu \nu}\,({\vec \rho}_{\mu}\,\times\,{\vec \rho}_{\nu})_3,
\label{photlag}
\ee
where $F_{\mu \nu} = \partial_{\mu}A_{\nu}-\partial_{\nu}A_{\mu}$, is the
field tensor for electromagnetic field
and $J^{\mu}$ is the hadronic part of the electromagnetic
current given by
\be
J^{\mu} = ({\vec \rho}_{\nu}\times{\vec \varrho^{\nu \mu}})_3 + (
{\vec \pi}\times(\partial^{\mu}\vec \pi+g_{\rho \pi \pi}{\vec \pi}\times{\vec
\rho}^{\mu}))_3,
\label{jmu}
\ee
with ${\vec \varrho_{\mu \nu}} = \partial_{\mu}{\vec \rho}_{\nu}-\partial_{\nu}
{\vec \rho}_{\mu}-g_{\rho \pi \pi}(\vec \rho_{\mu}\times\vec \rho_{\nu})$.
$\vec\pi$, $\vec\rho^\mu$ and $A^\mu$ represent the $\pi$, $\rho$ and photon
fields respectively and the arrows represent vectors in isospin space.
$g_{\rho\pi\pi}$ denotes the
coupling strength of the $\rho\pi\pi$ vertex, fixed from the
observed decay width $\rho\ra\pi\pi$.
We have also considered the photon
production due to the reactions $\pi\,\eta\,\rightarrow\,\pi\,\gamma$,
$\pi\,\pi\,\rightarrow\,\eta\,\gamma$ and the decay
$\omega\,\ra\,\pi\,\gamma$ using the interaction given in~\cite{GSW}
and vector meson dominance~\cite{sakurai}:
\bea
{\cal L} &=&
\frac{g_{\rho \rho \eta}}{m_{\eta}}\,
\epsilon_{\mu \nu \alpha \beta}\partial^{\mu}{\rho}^{\nu}\partial^{\alpha}
\rho^{\beta}\eta
+\frac{g_{\omega \rho \pi}}{m_{\pi}}\,
\epsilon_{\mu \nu \alpha \beta}\partial^{\mu}{\omega}^{\nu}\partial^{\alpha}
\rho^{\beta}\pi\nonumber\\
&&+\sum_{V=\rho,\omega}\frac{em_{V}^2}{g_{V}}V^{\mu}A_{\mu}
\label{etaro}
\eea
where $\epsilon_{\mu \nu \alpha \beta}$ is the
totally anti-symmetric Levi-Civita tensor.
The invariant amplitudes for all the reactions
are given in Ref.~\cite{17th}. The values of $g_{\rho\rho\eta}$ and
$g_{\omega\rho\pi}$ are fixed from the observed decays,
$\rho\,\ra\,\eta\,\gamma$
and $\omega\,\ra\,\pi\,\gamma$ respectively.
The constant $g_{V}$ is determined from the
decays, $V\ra\,e^+e^-$.
Photon production due to the process 
$\pi\,\rho\,\ra\,a_1\,\ra\,\pi\,\gamma$ is
also taken into consideration (for details of interaction
vertices see~\cite{15th}). 

In Fig.~\ref{fig2} the
static (fixed temperature) photon spectra is shown
for $T=180$ MeV. The solid line shows the enhanced
yield with in-medium masses and vacuum widths compared 
to the yield obtained with vacuum masses and widths
(dotted line).
The long-dashed line showing 
the result with in-medium width and vacuum masses
does not differ substantially from
the dotted line, because we find that the effects of 
the change in the decay width both on the phase space
and the production cross section are small.
We observe that the contributions from the
decays of baryonic resonances ($N(1520),\,N(1535),\,
N(1440),\,\Delta(1232),\,$ and $\Delta(1620)$) are also small
(filled circle). The values of the decay widths, $R\,\ra\,N\,\gamma$,
where $R$ and $N$ denote the baryonic resonances and the nucleon
respectively, are taken from the particle data book~\cite{PDG}. 
It is observed that the photon yield with universal
mass variation scenarios, I is almost an order
magnitude larger than the case II with large broadening
of the $\rho$. 
Reduction in the hadronic masses causes an enhancement 
in the thermal distribution
of mesons because of which the rate of thermal emission 
of photons increases. The scenario with
vacuum values of both masses and width (dotted line) does not
show any appreciable difference from (II).  

\bef
\centerline{\psfig{figure=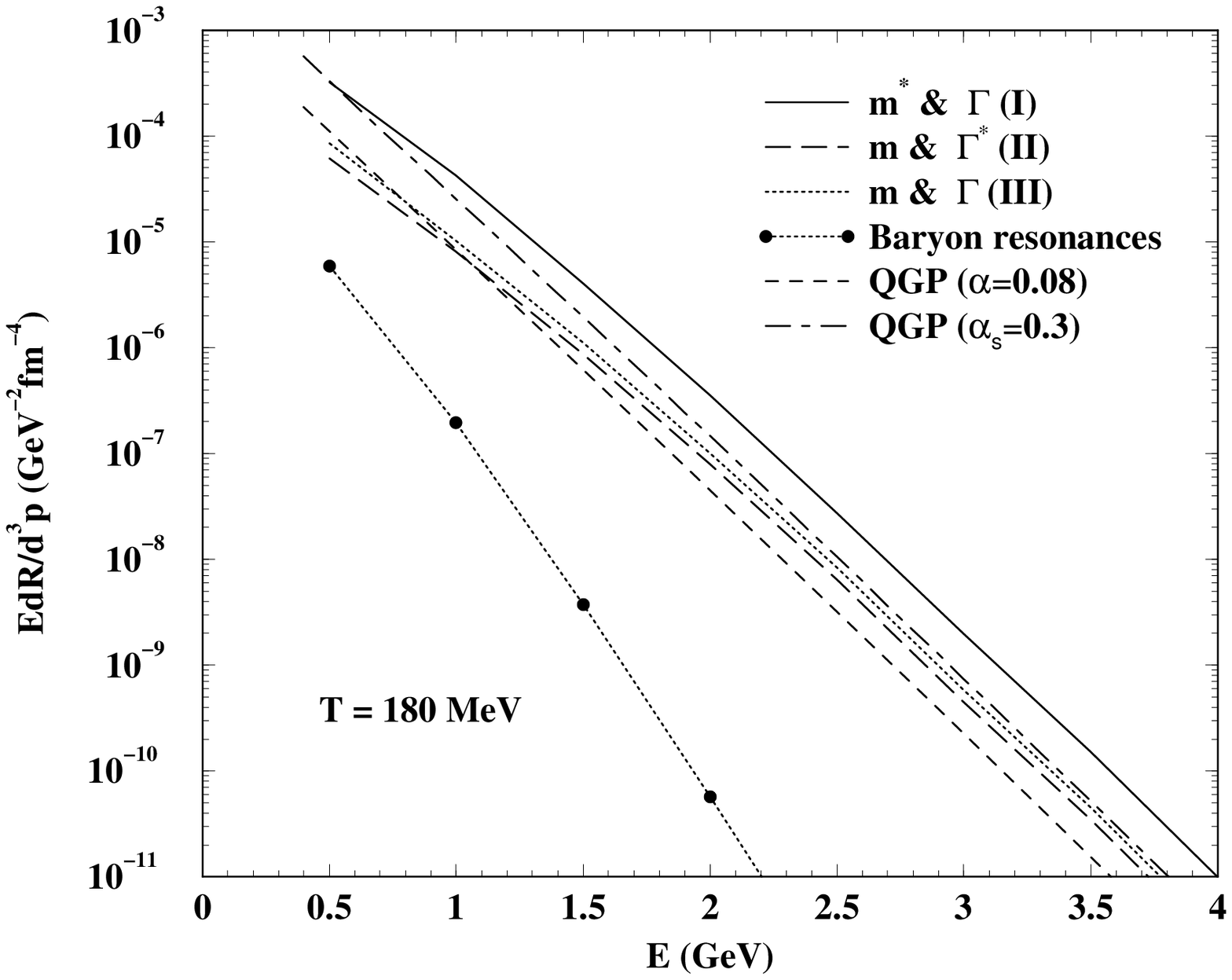,height=7.5cm,width=8cm}}
\caption{~The photon production rate as function of its energy
at a temperature, $T=180$ MeV. Solid (long dashed) line indicates 
rate for scenario I (scenario II).  Dotted line represents the spectra
without any medium effects. Filled circles indicate photons
originating from baryonic resonance decays. 
Short dashed (dot-dashed) line
indicates the photon emission rate from a thermalized
two-flavor quark gluon plasma for $\alpha_s=0.08 (0.2)$. 
}
\label{fig2}
\eef

For the sake of comparison we have also shown the photon emission
rate from a thermalized QGP. The photon spectra
from QGP includes Compton, annihilation, bremsstrahlung
and annihilation with scattering 
processes~\cite{16th,rbaier,paurenche,thoma} (see also \cite{dutta}).
Here we have mainly focused on photon production from hot hadronic matter
and hence do not consider emission rate from QGP beyond two 
loops~\cite{comm}.
From the lattice results~\cite{latt} the value
of $\alpha_s\sim 0.3 (g^2=4)$ at $T=180$ MeV.
At $T=180$ MeV 
the emission rates from the QGP (dot-dashed) and hot hadronic 
matter (scenario II and III) are comparable for $p_T>1.5$ GeV.
The photon yield from hadronic matter for scenario I 
is substantially larger than that from QGP. 
However, it should be mentioned here that
the photon emission rate from QGP 
is evaluated in Refs.~\cite{16th,rbaier,paurenche}
by using hard thermal loop (HTL) approximations, 
which is valid for the value of the color charge, $g<<1$.
The validity of the HTL approximation to higher
values of $g$ is doubtful. 
In fact, such an extrapolation in the value of $g$ (from 
$g<<1 $ to $g\,\sim 2$) introduces 
an uncertainty by a factor of $\sim 5$. Due to these uncertainties
we must be careful in making any firm conclusion
from the results obtained on the basis of the HTL approximations.
For $\alpha_s=0.08 (g=1)$ the 
emission rate from QGP (short-dashed line) 
is much smaller than the rates from hadronic matter.  

In Fig.~\ref{fig3} the $p_T$ distribution of photons (prompt+thermal)
is compared with the WA98 data.
Within the framework of the transport model
the data is well reproduced when the hadronic masses  
are allowed to vary according to the eqs.~\ref{emass}
(long-dash line, scenario I). 
However, when scenario (II) is considered for thermal photons (dash dotted line)
the experimentally observed ``excess'' photon 
in the region $1.5\le\,p_T$ (GeV) $\le 2.5$ 
is not reproduced. The dotted line indicates results
with vacuum masses and widths. 

\bef
\centerline{\psfig{figure=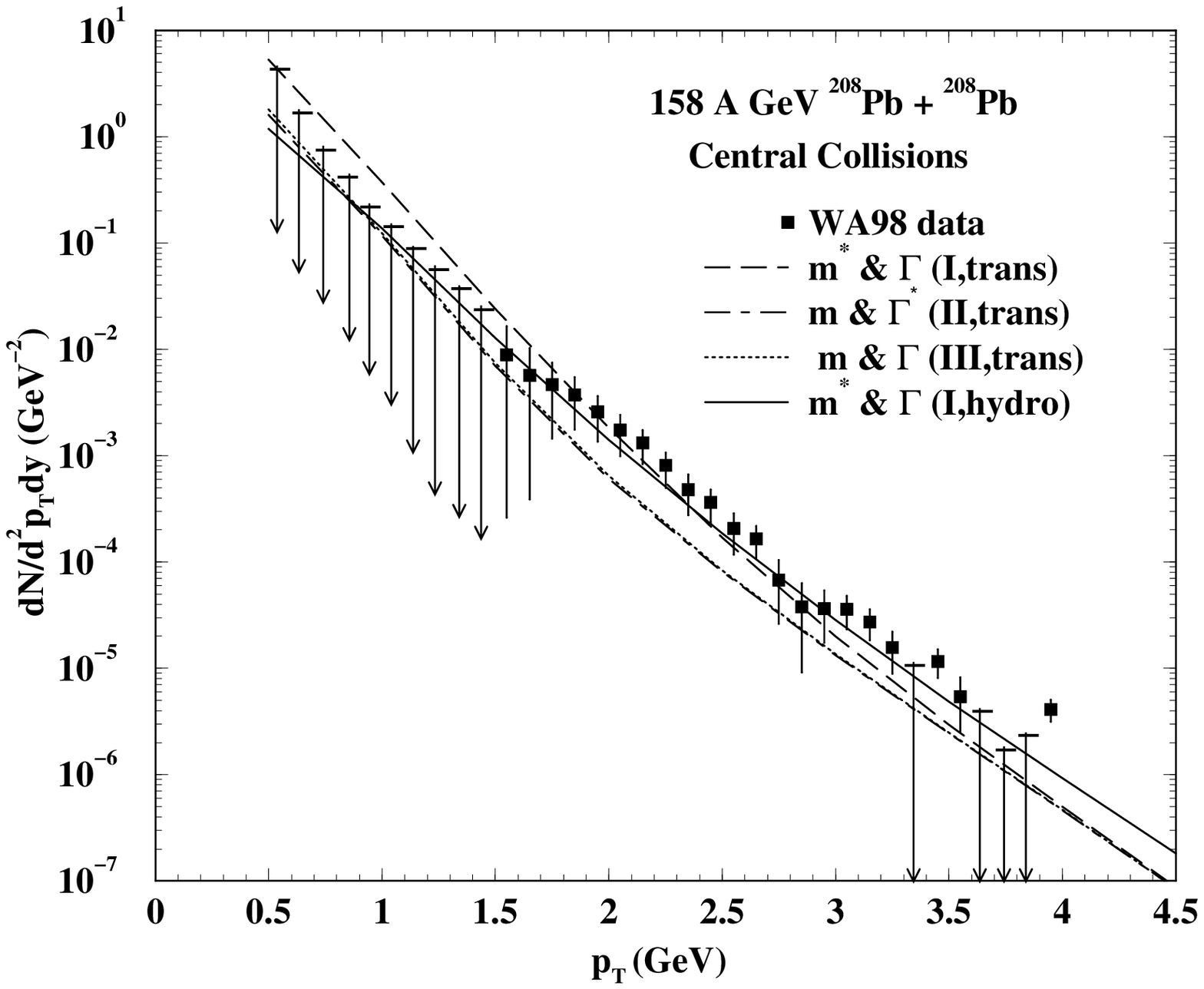,height=7.5cm,width=8cm}}
\caption{Total (prompt+thermal) photon yield in Pb + Pb collisions
at 158 A GeV at CERN-SPS. The theoretical
calculations contain hard QCD and thermal photons. 
The system is formed in the hadronic phase with initial temperature 
$T_i=200$ MeV; `trans' indicates the results 
for the cooling law~\protect\ref{cool}. 
}
\label{fig3}
\eef
 
The photon yield for the scenario (I) shown 
in Fig.~\ref{fig3} (solid line) is obtained by folding
the static rate with the space time history of the system governed 
by relativistic hydrodynamics
from the initial to the freeze-out state (freeze-out
temperature $\sim 120$ MeV); the data is
well reproduced in this case also. 

The agreement between the  $p_T$ spectra of photon obtained
with two different types of space time
evolution scenarios (transport model and hydrodynamics) can be explained 
as follows.
We find that the variation of temperature with time (cooling law)
in eq.~\ref{cool} is slower than the one obtained by solving 
hydrodynamic equations. As a consequence the thermal system
has a longer life time than the former case, allowing the
system to emit photons for a longer time. 
In case of hydrodynamics this is compensated by the transverse kick 
due to radial velocity of the 
expanding matter which 
shifts some of the photon multiplicity from lower values of transverse 
momentum towards the higher region.

\section*{7. Dilepton Spectra}

Now we study the invariant mass distribution 
of lepton pairs measured by CERES/NA45 collaboration
~\cite{ceres} in Pb + Au collisions. 
We consider the thermal dilepton production in the hadronic medium due to
the process $\pi^+\pi^-\,\ra\,e^+e^-$, 
known to be the most dominant
source of dilepton production from hadronic matter.
In the thermal system the width of the $\omega$ meson 
can be large due to various reactions occurring in the
thermal bath, the most dominant process, among others,
is $\omega\pi\,\leftrightarrow\,\pi\pi$~\cite{15th,unst}. 
As a consequence of this, the life time of the $\omega$
meson could become smaller compared to the life time 
of the hadronic system, enabling 
it to decay in the interior of the system. In view of this
the in-medium decay $\omega\,\ra\,e^+e^-$ is also taken 
into account. The required interaction vertices have been obtained
from eqs.~(\ref{photlag}), (\ref{jmu}) and (\ref{etaro}). 
In Fig.~\ref{fig5} the experimental data is 
compared with the theoretical results for $dN_{ch}/d\eta=270$. 
The effective masses and widths of $\rho$ and $\omega$
mesons appearing in the processes $\pi\pi\,\ra\,\rho\,\ra\,e^+e^-$
and $\omega\,\ra\,e^+e^-$ respectively 
are taken from Eq.~\ref{emass} (for details on the the
thermal emission rate of lepton pairs see Ref.~\cite{15th}).
Contributions from hadronic decays at freeze-out (background) 
is taken from the second of Ref.~\cite{ceres} (dotted line). 
The observed enhancement of the dilepton yield
around $M\sim 0.3 - 0.6 $ GeV can be reproduced
by the scenarios (I) (long-dashed) and (II) (dot-dashed) when the cooling
law is taken as~eq. \ref{cool}. The data is also well reproduced
when hydrodynamics is used to describe the space
time evolution of the system for the scenario I (solid line).
However, with
vacuum properties of the vector mesons the low mass
enhancement can not be reproduced (short-dashed), indicating the
change of vector meson properties in the medium. 
It is interesting to recall here that in a recent experiment
Ozawa {\it et al} ~\cite{ozawa} has observed a significant 
enhancement in the dilepton yield in p + Cu collisions as compared
to p + C collisions below the $\omega$ peak. 

\bef
\centerline{\psfig{figure=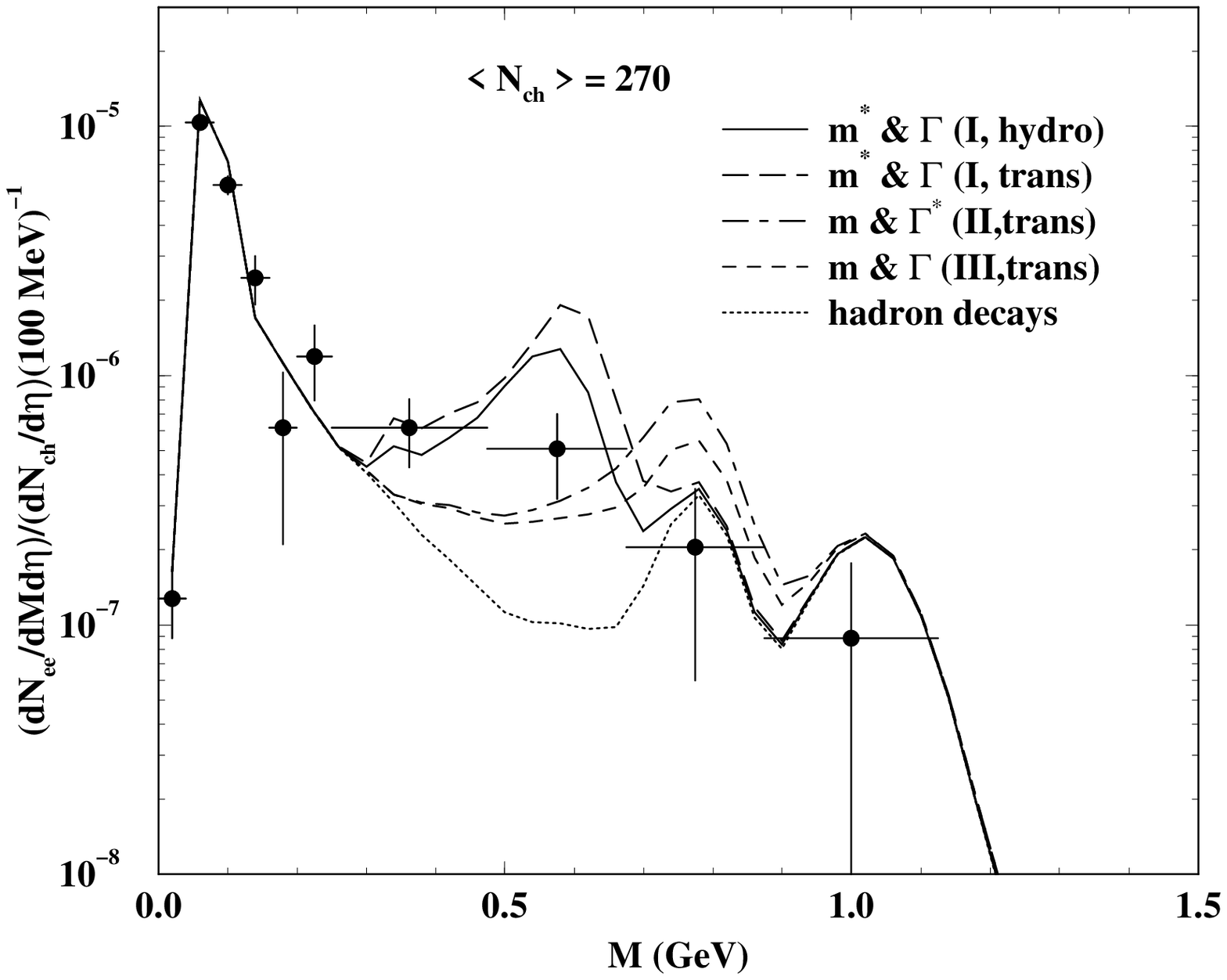,height=5.5cm,width=8cm}}
\caption{Dilepton spectra for $\lgl N_{ch}\rgl$=270 for
different scenarios as indicated in the text.
}
\label{fig5}
\eef

\section*{8. Summary and Discussions}

Let us now summarise the results. We reiterate that
our main contention in this work is to comment on the nature of medium effects
vis-a-vis mass shift as opposed to broadening of the spectral density
of hadrons. Such a study is only possible through the
electromagnetic probes since they are sensitive to the evolution of
the spectral function of vector mesons. The freeze-out conditions
within the model used for evaluating the photon and dilepton spectra
have been estimated through the NA49 pion and proton spectra.
Firstly, we find that the WA98 photon transverse momentum
spectra and the CERES dilepton low invariant mass spectra
is well explained 
if we assume a chirally restored initial phase at a temperature $\sim 200$
MeV where the masses of the hadrons tend to zero and grow to
their vacuum values with the decrease of temperature. 
In fact, close to the critical temperature for chiral symmetry
restoration (in our case~$\sim\,T_i= 200$ MeV) the description
of the system either in terms of partonic or hadronic degrees
of freedom become dual, although the non-perturbative effects
may still be important. For example, the emission rate of dileptons
from pion annihilation will resemble that from $q\bar{q}$ annihilation,
because of the complete extinction of the intermediary $\rho$ at the
critical temperature~\cite{14th,15th}. This indicates that the  $q\bar{q}$
interaction in the vector channel has become weak, signalling the onset of
chiral transition.  Whether such
a state is synonymous with deconfined matter (QGP) is still
an unsettled issue. A similar value of the initial temperature is 
obtained in Refs.~\cite{6th,7th,ss,20th}. 
Based on the present and our earlier results~\cite{6th}, 
it is fair to say that a simple hadronic model with vacuum properties
of hadrons is inadequate to explain the above experimental data.
Either a substantial change in the in-medium hadronic spectral
function or the formation of the QGP is required. 
More importantly, we find that the $p_T$ distribution
of photons changes significantly with a reduced mass scenario and
is almost unaffected by the broadening of the
vector meson spectral function in the medium.
The invariant mass distribution of the lepton pairs on the 
other hand can be explained with both a reduction in the mass as
well as with an enhanced width of the vector mesons.
Thus by looking only at the dilepton spectra 
it is difficult to differentiate the above scenarios; we need
to analyse both the photon and dilepton spectra simultaneously.
With better statistics we might succeed in ruling out or at
least restricting one of the scenarios. We expect that such
a situation should be realized at
the Relativistic Heavy Ion Collider (RHIC).

\noindent{{\bf Acknowledgement}: 
We are grateful to Tetsuo Hatsuda for useful
comments on this manuscript. Useful discussions 
with Tapan K. Nayak is thankfully acknowledged.  
J.A. is grateful to the Japan
Society for Promotion of Science (JSPS) for financial support.
J.A. is also supported by Grant-in-aid for Scientific
Research No. 98360 of JSPS.


\begin{thebibliography}{99}
\bibitem{comm} The thermal photon emission rate from QGP complete to 
order $\alpha_s$
has recently been obtained in~\cite{pmy} which was used by 
Huovinen et. al.~\cite{hrr} to analyze the WA98 photon data).
\end{thebibliography}
\end{document}